\documentstyle[12pt,epsf]{article}
\makeatletter

\@addtoreset{equation}{section}
\makeatother

\newcommand{\lsim}{\mathop{}_{\textstyle \sim}^{\textstyle <}}
\begin{document}
\begin{titlepage}
\vspace*{-10mm}
\begin{flushright}
{\small 
KANAZAWA-02-28\\
KUNS-1812\\
NIIG-DP-02-7\\
TU-674\\[-1mm]
hep-ph/0211347}
\end{flushright}
\begin{center}
\vspace*{6mm} 
    
{\Large\bf Flavor violation in supersymmetric theories\\[2mm]
with gauged flavor symmetries}
\vspace*{8mm}

Tatsuo~{\sc Kobayashi},\footnote{E-mail: 
kobayash@gauge.scphys.kyoto-u.ac.jp}
Hiroaki~{\sc Nakano},\footnote{E-mail: 
nakano@muse.sc.niigata-u.ac.jp}
Haruhiko~{\sc Terao},\footnote{E-mail: 
terao@hep.s.kanazawa-u.ac.jp}
and
Koichi~{\sc Yoshioka}\footnote{E-mail: 
yoshioka@tuhep.phys.tohoku.ac.jp}
\vspace*{4mm}
  
$^*${\it Department of Physics, Kyoto University, Kyoto 606-8502,
Japan}\\[1mm]
$^\dagger${\it Department of Physics, Niigata University, Niigata 
950-2181, Japan}\\[1mm]
$^\ddagger${\it Institute for Theoretical Physics, Kanazawa
University, Kanazawa 920-1192, Japan}\\[1mm]
$^\S${\it Department of Physics, Tohoku University, Sendai 980-8578, 
Japan}
\vspace{8mm}
  
\begin{abstract} \noindent
In this paper we study flavor violation in supersymmetric models with
gauged flavor symmetries. There are several sources of flavor
violation in these theories. The dominant flavor violation is the
tree-level $D$-term contribution to scalar masses generated by flavor
symmetry breaking. We present a new approach for suppressing this
phenomenologically dangerous effects by separating the flavor-breaking
sector from supersymmetry-breaking one. The separation can be achieved
in geometrical setups or in a dynamical way. We also point out that
radiative corrections from the gauginos of gauged flavor symmetries
give sizable generation-dependent masses of scalars. The gaugino mass
effects are generic and not suppressed even if the dominant $D$-term
contribution is suppressed. We also analyze the constraints on the
flavor symmetry sector from these flavor-violating corrections.
\end{abstract}
\end{center}
\end{titlepage}
\setcounter{footnote}{0}

\section{Introduction}

Supersymmetric extension of the standard model (SM) has been found to
be very attractive, in particular, as a solution to the hierarchy
problem. Superpartners of the SM fields are expected to be detected in
future experiments. Even at present, supersymmetry (SUSY) breaking
parameters are constrained from flavor-changing neutral current (FCNC)
processes as well as CP violation~\cite{FCNC}. That is the so-called
SUSY flavor problem and requires sfermion masses between the first and
second generations being degenerate, unless they are sufficiently
heavy or fermion and sfermion mass matrices are aligned quite
well. Such requirements for sfermion masses have been tried to be
realized by considering flavor-blind SUSY breaking and/or mediation
mechanisms. Actually, various types of flavor-blind mechanisms have
been proposed in the literature~\cite{gravity}-\cite{others}.

Understanding the origin of fermion masses and mixing angles is also
one of the important issues in particle physics. Three copies of SM
generations have the exactly same quantum numbers except for their
masses, i.e.\ Yukawa couplings to the Higgs field. In the SM, the
exact flavor universality is violated only in the Yukawa sector. One
expects that the hierarchical structure of Yukawa matrices is
explained by some dynamics beyond the SM, and such additional dynamics
necessarily leave some imprint of flavor violation. In supersymmetric
models, a mechanism for realistic fermion masses breaks the flavor
universality and generally makes the corresponding sfermion masses
flavor-dependent.

One of the salient mechanisms for generating hierarchical Yukawa
couplings is the Froggatt-Nielsen (FN) mechanism with additional
$U(1)$ gauge symmetries~\cite{FN,IR}. In the FN scenarios,
flavor-dependent $U(1)$ charges are assigned to matter fields so that
realistic Yukawa matrices are effectively realized in terms of
higher-dimensional operators. There is a certain reason to believe
that the $U(1)$ symmetries should be gauged; any global symmetry is
expected to be unstable against quantum gravity and hence
accidental. Therefore the $U(1)$ flavor symmetries that exactly
control such operators should be gauged. Consequently, after flavor
symmetry breaking, the auxiliary $D$ fields of the $U(1)$ vector
multiplets give additional contribution to sfermion masses, which is
proportional to $U(1)$ charges and hence
flavor-dependent~\cite{anom1,anom2}. That indeed gives significant
modification of sparticle spectrum and in some cases is confronted
with the SUSY flavor problem. This is called the $D$ problem in the
present paper.

In this paper, we present an idea for suppressing the flavor-dependent 
$D$-term contribution to sfermion masses. The $D$-term contribution is 
generated when one integrates out heavy fields which develop vacuum
expectation values (VEVs) that break the $U(1)$ gauge symmetries. As
will be reviewed below, the modification of low-energy spectrum is
determined by soft SUSY-breaking masses of the heavy fields. Our idea
is that the $D$-term contribution is suppressed if the soft masses of
the heavy fields can be made small. We will present illustrative
models where this idea is realized in a dynamical or geometrical
way. Unlike other approaches, the models presented here have an
advantage that origins of the Yukawa hierarchy can be addressed within
the same framework.

However, even if the dominant $D$-term contribution is suppressed,
there still remains a flavor-violating effect from the $U(1)$ gauge
symmetries; flavor-dependent scalar masses are radiatively generated
by $U(1)$ gaugino loop graphs, which involve the
gaugino-fermion-sfermion vertices proportional to $U(1)$ quantum
numbers. Such radiative effect is described in terms of
renormalization-group equations (RGEs) for scalar masses. Unlike the
above tree-level $D$-term contribution, the gaugino loop effect does
not depend on SUSY-breaking scalar masses. Therefore even with reduced
$D$-term contribution, e.g.\ by assuming the universal sparticle
spectrum, flavor violation from the $U(1)$ flavor symmetries still
remains and turns to be detectable signatures for flavor physics.

This paper is organized as follows. In Section 2 we describe scalar
masses in supersymmetric theories with a gauged abelian flavor
symmetry. In addition to the $D$-term contribution, we point out
subleading but sizable flavor violation due to the abelian gaugino. In
Section 3, we will discuss how to suppress the dominant $D$-term
flavor violation. After a brief survey of the existing proposals in
Section 3.1, we present in Section 3.2 the models with extra
dimensions, while another model based on four-dimensional
superconformal dynamics is presented in Section 3.3. We also examine
how much degeneracy of sfermion masses is expected in these models. In
Section 4, radiative corrections from the soft gaugino mass are shown
to be potentially dangerous and give significant constraints on model
parameters such as soft gaugino masses. Section 5 summarizes our results.

\section{Sfermion masses with $U(1)$ gauge symmetry}

\subsection{$D$-term contribution}

In this section, we discuss sfermion masses in the presence of a
$U(1)$ horizontal gauge symmetry, denoted by $U(1)_X$ throughout this
paper. The $D$-term effect has been considered in various
contexts~\cite{otherD}. As simple examples, we study two types of
models where a non-vanishing VEV of the auxiliary component $D$ of
$U(1)_X$ vector multiplet is actually generated. However the
properties presented here are generic for any model of $D$-term
contribution.

As the first example, let us consider a pseudo-anomalous abelian gauge
symmetry, which often appears in string models~\cite{stringU1}. In
this case, the Fayet-Iliopoulos (FI) term is generated by a
non-vanishing VEV of the dilaton or moduli field, whose nonlinear
shift cancels the $U(1)_X$ anomaly. We treat the coefficient of the FI
term as a constant, for simplicity.\footnote{If the dilaton and moduli
fields are treated as dynamical fields, the FI term and K\"ahler
metric of the FN field depend on these fields. In this case, as was
shown in Ref.~\cite{anom2}, the formula for $D$-term contribution
becomes different from that in softly-broken global SUSY models with a
constant FI term. The suppression mechanism presented in this paper
should be reconsidered in such a case.} 
The supersymmetric scalar potential relevant to the $U(1)_X$ gauge
sector is written as
\begin{equation}
  V_{\rm SUSY} \;=\; -\frac{1}{2g_X^2}D^2  +D\bigg(\xi_{\rm FI}
  +\sum_{i=1}^N q_i |\phi_i|^2 +\sum_{j=1}^{\bar N} \bar q_j
  |\bar\phi_j|^2 \bigg)+\cdots.
  \label{Vsusy}
\end{equation}
Here $g_X$ is the gauge coupling and $\xi_{\rm FI}$ is the coefficient
of the FI term, which we take positive without lose of generality. The
$U(1)_X$ charges of scalars $\phi_i$ and $\bar\phi_i$ are denoted as
$q_i$ $(>0)$ and $\bar q_i$ $(<0)$, respectively. The equation of
motion for $D$ is
\begin{equation}
  \frac{D}{g_X^2} \;=\; \xi_{\rm FI} +\sum_{i=1}^N q_i |\phi_i|^2 
  +\sum_{j=1}^{\bar N} \bar q_j |\bar\phi_j|^2.
  \label{Deom}
\end{equation}
Then in the supersymmetric limit $D=0$, the negatively-charged fields
$\bar\phi_i$ generally develop nonzero VEVs and the abelian gauge
symmetry is broken at $M_X$ $\sim \xi_{\rm FI}^{1/2}$. The scalar
fields with vanishing VEVs remain massless but $\bar\phi_i$ decouple
around the $M_X$ scale.

In this work, we assume that the $U(1)_X$ breaking scale 
$M_X$ $\sim \xi_{\rm FI}^{1/2}$ is smaller than a cutoff $\Lambda$ of
the theory. It is known~\cite{FN,IR} that the ratio 
$\xi_{{\rm FI}}^{1/2}/\Lambda$ can be an origin of the hierarchy of
Yukawa couplings. Consider, for example, a superpotential $W$ includes
the following non-renormalizable operators
\begin{equation}
  W \;=\; y_{ij}\bigg(\frac{\bar\phi}{\Lambda}\bigg)^{n_{ij}}
  \phi_i\phi_j H,
  \label{FNspot}
\end{equation}
where $H$ denotes the electroweak Higgs field, and $\bar\phi$ is a
negatively-charged field that develops a nonzero VEV of order 
$\xi_{\rm FI}^{1/2}$. The power $n_{ij}$ is determined by the $U(1)_X$ 
charge conservation to be $n_{ij}=(q_i+q_j+q_H)/|\bar{q}|$. The
operators (\ref{FNspot}) induce the effective Yukawa couplings
\begin{equation}
  y'_{ij} \;=\; \lambda^{n_{ij}} y_{ij}\,, \qquad 
  \lambda\equiv
  \bigg(\frac{\langle\bar\phi\rangle}{\Lambda}\bigg)^{|\bar{q}|}.
\end{equation}
The factor $\lambda$ represents a unit of hierarchy of Yukawa
couplings and is usually taken to be of the order of the Cabibbo
angle. In this way, a realistic hierarchy of low-energy Yukawa
couplings can be obtained by assigning different charges to matter
fields $\phi$~\cite{FN,IR}. 

We add a remark on possible realization of the FN mechanism in
weakly-coupled heterotic string models. In this case, the $U(1)$
anomaly is cancelled by a nonlinear shift of the dilaton-axion
multiplet~\cite{GS}, and the FI term $\xi_{{\rm FI}}$ is generated at
loop level. Consequently $\xi_{{\rm FI}}$ naturally has an appropriate
size for the Yukawa hierarchy. This possibility provides us with a
strong motivation for regarding an anomalous $U(1)$ as a gauged flavor
symmetry. Note also that in the case of anomalous $U(1)$ symmetry, the
axion-gauge mixing generates an additional contribution to the gauge
boson mass, $2(\xi_{\rm FI}/\Lambda)^2$. Since this contribution is
suppressed by $\xi_{\rm FI}/\Lambda^2$ compared with that from the
scalar VEV $\langle\bar\phi\rangle$, we will neglect this effect in
our analysis below.

The SUSY vacuum is shifted when soft SUSY-breaking masses for scalars
are introduced. The scalar potential is now given by
\begin{equation}
  V \;=\; V_{\rm SUSY} + \sum_{i=1}^N m_i^2 |\phi_i|^2 
  +\sum_{j=1}^{\bar N} \bar m_j^2 |\bar\phi_j|^2 \,+\cdots,
  \label{V}
\end{equation}
where $m_i^2$ and $\bar m_j^2$ are arbitrary mass parameters. The
ellipsis denotes other SUSY-breaking terms irrelevant to our
analysis. In $V_{\rm SUSY}$, the $D$ component is replaced with scalar
fields through the equation of motion (\ref{Deom}). Minimizing the
potential with respect to negatively-charged fields, we find that the
$D$ component obtains a VEV
\begin{equation}
  \langle D\rangle \;=\; \frac{\bar m^2}{|\bar q|}.
  \label{Dvev}
\end{equation}
In this expression, $\bar m^2/|\bar q|$ is the minimum value of 
$\bar m^2_i/|\bar q_i|$ in the model, since the effective potential
around the minimum along the $\bar\phi_i$ direction takes a value of
$O(\bar m^2_i/|\bar q_i|)\,\xi_{\rm FI}$. Without superpotential, only
the scalar fields with such minimal value of ratio, 
$\bar m^2/|\bar q|$, can contribute to the above equation. For
example, in the case of universal scalar masses, (a combination of)
scalar fields with the largest negative value of $U(1)$ charge obtain
the VEV (\ref{Dvev}). In this way, the $D$-flat direction is lifted by
SUSY-breaking masses of scalar fields with negative quantum
numbers. From Eqs.~(\ref{Vsusy}) and (\ref{Dvev}), one finds a formula
for the $D$-term contribution to the masses of light scalars:
\begin{equation}
  m^2_{D_i} \;=\; q_i\langle D\rangle \;=\; 
  \frac{q_i}{|\bar q|}\,\bar m^2.
  \label{Dcont}
\end{equation}

The most important property of the formula (\ref{Dcont}) is that the 
induced scalar masses squared are proportional to their $U(1)_X$
charges. This fact gives a significant implication to flavor
physics. As explained above, realistic low-energy Yukawa couplings are
generated by assigning different charges to matter fields. If all
SUSY-breaking masses are of the same order of magnitude, this scalar
mass difference leads to large FCNC amplitudes. For example,
lepton-flavor violation from flavor symmetry $D$-terms was discussed
in~\cite{DLFV}.

Besides the flavor problems, the induced scalar masses (\ref{Dcont})
have several interesting properties. Firstly, the contribution is
independent of the $U(1)_X$ gauge coupling constant $g_X$. Therefore
the formula (\ref{Dcont}) remains valid even if $g_X$ is small. [Note,
however, that the complete global limit ($g_X\to 0$) cannot be taken
since the $U(1)_X$ symmetry is broken only if the condition
$g_X^2>\bar m^2/(-\bar q\xi_{\rm FI})$ is satisfied so that the scalar
potential (\ref{V}) is unstable around the origin of the moduli space.] 
Secondly, the $D$-term contribution does not depend on the
symmetry-breaking scale $M_X$, either. The $D$-term contribution is
proportional to a tiny deviation from supersymmetric conditions (flat
directions), and the deviation is determined by supersymmetry breaking
independently of the gauge symmetry breaking~\cite{otherD}. As a result
of these properties, the scalar masses induced via the $D$ term appear
for any value of gauge coupling and symmetry-breaking
scale\rlap.\,\footnote{This is true provided that soft SUSY-breaking
terms are present already at the scale of scalar VEVs. On the other
hand, it will be intuitively clear that the $D$-term contribution is
absent if soft masses arise only at low energy as in the gauge
mediation of SUSY breaking~\cite{GM}. This can be seen from the fact
that, to correctly determine the scalar VEVs, one has to use the
renormalization-group improved potential~\cite{RGpot} in which all
running parameters are evaluated at the VEV scale.} It is also found
from (\ref{Dcont}) that the normalization of $U(1)_X$ charges does not
affect the size of $D$-term contribution.

The $D$-term contribution also appears for non-anomalous abelian
gauge symmetries. Here we consider a model which contains vector-like
fields $Y$ and $\bar Y$ with $U(1)_X$ charges $\pm q_Y$, and a gauge
singlet $Z$. Taking a renormalizable and gauge-invariant superpotential
\begin{equation}
  W \;=\; f Z(Y\bar Y - M_X^2),
  \label{superpot}
\end{equation}
and introducing soft SUSY-breaking terms, we obtain at the minimum of
the scalar potential a VEV
\begin{equation}
  |\langle Y\rangle|^2-|\langle\bar Y\rangle|^2 \;\simeq\; 
  \frac{1}{2q_Y^2g_X^2}\big(m_{\bar Y}^2-m_Y^2\big),
  \label{diff}
\end{equation}
where $m_Y^2$ and $m_{\bar Y}^2$ are the soft scalar masses of $Y$ and
$\bar Y$, respectively. If one introduces an $R$ symmetry under which
the singlet $Z$ has charge $+2$, the superpotential (\ref{superpot})
is the most generic one. The abelian gauge symmetry is broken at 
$\langle Y\rangle\simeq\langle\bar Y\rangle\simeq M_X$. The result
(\ref{diff}) is not modified even if one stabilizes scalar fields with
other types of superpotentials. When the theory contains a chiral
multiplet with a charge $q_i$, the induced mass for its scalar
component is given by
\begin{equation}
  m_{D_i}^2 \;=\; q_iq_Yg_X^2\big(|\langle Y\rangle|^2
  -|\langle\bar Y\rangle|^2\big) \;\simeq\;
  \frac{q_i}{2q_Y}\big(m_{\bar Y}^2-m_Y^2\big).
\end{equation}
This has the same form as in Eq.~(\ref{Dcont}) and shares the
properties discussed above with the anomalous $U(1)$ case. Note that
in this model the $D$-term contribution vanishes as long as soft
scalar masses are universal. However unless the exact universality is
a result of symmetries of models, the degenerate spectrum is expected
to be split by radiative corrections governed by renormalization-group
(RG) evolutions. Then non-vanishing $D$-terms are generated.

In this way, the $D$-term contribution generically appears in
supersymmetric models with abelian gauge symmetries. We note that
abelian gauge factors always appear when the rank of gauge group
is reduced through gauge symmetry breaking, e.g.\ as in grand unified
theories (GUTs). For example, there are several proposals for realistic
Yukawa matrices where matter multiplets of different generations
belong to different representations of GUT gauge
group~\cite{nu}. After gauge symmetry breaking, non-vanishing VEVs of 
abelian $D$ components induce violation of flavor universality. One
important notice in this case is that the $U(1)_X$ gauge couplings
cannot be arbitrarily small because they are unified into GUT gauge
group which also contains the SM ones.

\subsection{scalar soft masses}

In this work, we focus on the effects of $U(1)_X$ gauge symmetry on
the scalars with the same quantum numbers of the SM gauge symmetry.
We neglect effects from Yukawa interactions. This is justified for the
light generations, for which the experimental constraints are severe.
When one includes the third generation with large Yukawa couplings
(e.g.\ for top quark), flavor violation generated by these Yukawa
couplings may be large. Experimental constraints are, however, still
weak for the third generation and will be a target of next generation
of experiments. We leave it to future investigations.

The scalar mass under consideration generally takes the form
\begin{equation}
  m_i^2 \;=\; m_{0_i}^2 + m_g^2 + m_{D_i}^2 + m_{X_i}^2
  \label{softmass}
\end{equation}
at the flavor symmetry breaking scale $M_X$. The first term on the
right-handed side is an initial value of the scalar mass. This part is
expected to be generated by SUSY-breaking dynamics and its structure
is highly model-dependent. To focus on the $U(1)_X$ part, we take an
assumption that $m_{0_i}^2$ are not dominant sources of flavor
violation. In what follows, $m_{0_i}^2$ are taken to be
generation-independent for simplicity, but this assumption is not
necessary in our concrete models in Section 3.2.

The second term $m_g^2$ contains the effects from all flavor-blind
gauge interactions such as GUT or SM ones. For example, the RG
evolution due to soft gaugino masses is described as
\begin{equation}
  m_g^2 \;=\; \frac{-8}{16\pi^2}\sum_a C^a_2(R) \int^{M_X}_\Lambda
  \frac{d\mu}{\mu}\, g_a^2 M_{\lambda^a}^2 +\cdots,
\end{equation}
where $g_a$ and $M_{\lambda^a}$ are the gauge coupling constant and
soft gaugino mass of the $G_a$ gauge sector, respectively. The scale
$\mu$ denotes the renormalization point and $C^a_2(R)$ is the
quadratic Casimir of $G_a$ for the corresponding scalar in the
representation $R$. In addition, there is another type of gauge
contribution if the theory includes a generation-independent abelian
factor like the SM hypercharge. That introduces $D$-term contribution
and the Tr($Qm^2$) term in scalar mass RGEs. It is potentially
important that low-energy sparticle spectrum is modified in the
presence of such extra $U(1)$ symmetries. Generation-blind but
intergeneration-dependent $D$-term contribution has been investigated
in various contexts~\cite{otherD}. In RG evolutions of scalar masses,
the net effect is to shift the low-energy value of hypercharge $D$
term which is proportional to Tr($Q_Ym^2$). All of these contributions
just scale overall magnitude of soft scalar masses and therefore are
harmless to the flavor problems.

The remaining two factors, the third and last terms in
(\ref{softmass}), are possible flavor-violating terms associated with
the $U(1)_X$ gauge symmetry. The form of $m_{D_i}^2$ was given in the
previous section,
\begin{equation}
  m_{D_i}^2 \;=\; q_i\langle D\rangle.
  \label{mD}
\end{equation}
The VEV $\langle D\rangle$ is written in terms of scalar VEVs which
break the flavor gauge symmetry, and the scalar VEVs are determined by
their soft masses. On the other hand, the last term represents
radiative corrections via RGEs from the $U(1)_X$ gauge interactions
and is explicitly given by
\begin{equation}
  \hspace*{1cm} 
  m_{X_i}^2 \;=\; \frac{-8q^2_i}{16\pi^2} \int^{M_X}_\Lambda
  \frac{d\mu}{\mu}\, g_X^2 M_{\lambda_X}^2 + \frac{2q_i}{16\pi^2}
  \int^{M_X}_\Lambda \frac{d\mu}{\mu}\, g_X^2 {\rm Tr} (Q_X m^2)
  +\cdots,
  \label{mX}
\end{equation}
where $M_{\lambda_X}$ is the $U(1)_X$ soft gaugino mass, $Q_X$ is the
charge operator, and the trace is taken over all scalar fields charged
under the $U(1)_X$ symmetry. The second term in (\ref{mX}) is
generally non-vanishing for the case of non-universal soft scalar
masses and also for anomalous $U(1)_X$ case even with universal scalar
masses. However it should be noted that the masses squared 
$\bar m_i^2$ of symmetry-breaking scalars $\bar\phi_i$ are also
affected by the same factor ${\rm Tr}(Q_Xm^2)$ in RG evolutions. That
is, $\bar m_i^2$ evaluated at the scale $M_X$ contains a term
\begin{equation}
  \bar m_i^2 \;=\; \cdots + \frac{2\bar q_i}{16\pi^2}
  \int^{M_X}_\Lambda \frac{d\mu}{\mu}\, g_X^2 {\rm Tr} (Q_X m^2).
\end{equation}
Interestingly, when substituted into the formula (\ref{Dcont}), this 
contribution exactly cancels the second term in (\ref{mX}). The same
is true for non-anomalous $U(1)$ cases, (2.10). Therefore we can
safely drop this type of effects in the following analysis and
concentrate on the first term in (\ref{mX}).

\medskip

We here comment on characteristic features of two flavor-violating
contributions from abelian gauge dynamics, $m_{D_i}^2$ and
$m_{X_i}^2$. They have rather different properties from each
other. The $D$-term contribution $m_{D_i}^2$ is the dominant,
tree-level source of flavor violation from the gauged flavor
symmetry. The induced masses squared of light scalars are linearly
proportional to their $U(1)_X$ charges, but independent of the gauge
coupling strength and the symmetry-breaking scale. On the other hand,
$m_{X_i}^2$ is the gaugino radiative correction and depends on the
values of gauge coupling and symmetry-breaking scale. As a result,
although loop-suppressed, this sub-dominant contribution can be
enhanced and become comparable with $m_{D_i}^2$. Moreover, $m_{X_i}^2$
does not depend on the magnitude of soft scalar masses, unlike 
$m_{D_i}^2$. This fact implies that even if the apparently dominant
contribution $m_{D_i}^2$ were suppressed, for example, by taking the
universality assumption, $m_{X_i}^2$ is left unchanged and becomes the
main source of flavor violation. Combining these two properties of
$m_{X_i}^2$, we find that the $U(1)_X$ gaugino effect is an
interesting and unexplored effect in supersymmetric models with flavor
$U(1)_X$ symmetries. This issue will be discussed in Section 4.

\section{Suppressing $D$-term contributions}

The aim of this section is to discuss how to suppress the tree-level
$D$-term contribution which is known to be a dominant source of flavor
violation with $U(1)$ flavor symmetry. We first mention several
possible approaches to the $D$ problem, some of which have been
discussed in the literature. Our concrete models will be presented in
Sections 3.2 and 3.3.

\subsection{Approaches to the D problem}

A well-known solution to the SUSY flavor problems is to take 
$m_{D_i}^2$ very large, i.e.\ of the order of $O(10)$
TeV~\cite{decoupling}. With such heavy scalars, flavor- and
CP-violating processes involving these fields are suppressed by their
large masses and do not lead to any severe constraints on
SUSY-breaking parameters. It is also interesting that together with
the FN mechanism, a larger $U(1)_X$ charge simultaneously leads to a
larger scalar mass and a smaller Yukawa coupling. This is precisely
the situation relevant to the first and second generations on which
the experimental constraints are severer. This elegant solution,
however, suffers from some unsatisfactory points, including the
destabilization of the true electroweak vacuum at two-loop
level~\cite{ccb}, fine-tuning of parameters required for the
CP-violation constraints~\cite{K}, etc. This solution also rules out
supersymmetry as a possible explanation of the experimental result for
the anomalous magnetic moment of the muon.

The $D$ problem becomes less severe if one assigns the same $U(1)_X$ 
charge to the matter fields in the same representation under the SM
gauge symmetry. It was pointed out that universal charges for
three-generation left-handed leptons can explain the large flavor
mixing observed at the neutrino experiments (but with a bit large
value of $U_{e3}$ matrix element~\cite{anarchy}). Within the FN
mechanism, however, this could only be applied to the lepton
sector. Mass hierarchy and small flavor mixing in the quark sector are
realized by assigning different $U(1)_X$ quantum numbers.

Another possibility is a cancellation of $m_{D_i}^2$ with other
contributions to scalar masses. An obvious choice comes from
superpotential $F$-terms. It has been argued that $F$- and $D$-term
effects add up to zero in specific situations~\cite{anom1,anom2}. A
cancellation may occur in $D$-term itself if there are several scalar
fields developing non-vanishing VEVs. However cancellation of more than 
one flavor-violating effects seems unnatural unless it follows from
some underlying mechanism.

Now, there is a simple and natural possibility which, to our
knowledge, has not been addressed in the literature. As was seen in
the previous section, the $D$-term contribution is proportional to a
deviation from the $D$-flat direction, and such deviation is
determined by the soft masses of the scalars that acquire flavor
symmetry-breaking VEVs. Accordingly, if these soft masses can be
reduced, the $D$-flat direction is not lifted and $D$-term
contributions become negligible. It is this possibility that we shall
pursue in the following part of this section. We will present two
illustrative examples where our idea can be realized. The point is to
separate the $U(1)_X$-breaking sector from the SUSY-breaking
sector. We construct toy models in which this separation is achieved
in a geometrical or dynamical way. In what follows, we assume, for
simplicity of presentations, that only a single field $\chi$ obtains a
VEV of the $U(1)_X$ flavor symmetry breaking.

\subsection{Suppression via extra dimensions}

We now describe how our idea of suppressing $D$-term contributions can
be realized by considering a suitable field configuration in 
higher-dimensional spacetime. For definiteness, we consider a 
five-dimensional theory and suppose that there are two
four-dimensional branes at different positions in the fifth dimension;
the visible-sector matter fields, i.e.\ quarks and leptons, are
assumed to be confined on one brane, while supersymmetry breaking
occurs on another brane (called the hidden brane). As was pointed out
in Ref.~\cite{RS}, this setup naturally explains the absence of 
uncontrollable gravity-mediated effects including large
flavor-violating ones, provided that there is no light mode whose
wavelength is longer than the distance between the two branes. This is
one of the most attractive points of considering the presence of
additional dimensions. However the $D$-term contribution to scalar
masses can be added~\cite{DAM}. If flavor $U(1)_X$ gauge symmetries
are introduced in this framework, suppressing their $D$ terms is
required.

We apply the idea of geometrical splitting to separate the flavor
symmetry-breaking sector from SUSY-breaking one. Specifically we
require that, in addition to the quarks and leptons, the $\chi$ field
is also confined to the visible brane. This is a setup crucial for our
scenarios. We further suppose that the $U(1)_X$ vector multiplet is
also confined on the visible brane. This is just a technical
assumption for simplifying discussion. If $U(1)_X$ is put in the bulk,
the supersymmetric five-dimensional abelian gauge theory has a rich
structure of vacua, which requires complicated analyses. We do not
pursue this possibility in this paper. On the other hand, there are
two choices for the SM gauge sector; on the visible brane or in the
bulk.

First let us consider the SM gauge multiplets on the visible brane. In
this case, all the SM fields and $\chi$ are on the visible
brane. Though couplings to the hidden brane are suppressed, soft terms
do not completely vanish. It has been shown that SUSY-breaking effects
are transmitted at loop level to the visible sector via superconformal
anomaly~\cite{AM}. The anomaly-mediated spectrum is roughly given by
\begin{eqnarray}
\begin{array}{ccl} 
  M_{\lambda_{\rm SM}} \;\sim\; 
  \displaystyle{\frac{g_{\rm SM}^2}{16\pi^2}m_{3/2}},
  && m_{\rm SM}^2 \;\sim\; \bigg(
  \displaystyle{\frac{g_{\rm SM}^2}{16\pi^2}m_{3/2}}\bigg)^2, \\[4mm] 
  M_{\lambda_X} \;\sim\; \displaystyle{\frac{g_X^2}{16\pi^2}m_{3/2}},
  && m_\chi^2 \;\sim\; 
  \bigg(\displaystyle{\frac{g_X^2}{16\pi^2}m_{3/2}}\bigg)^2,
  \label{3AM}
\end{array}
\end{eqnarray}
where $M_{\lambda_{\rm SM}}$, $m_{\rm SM}^2$ and $g_{\rm SM}$ are the
gaugino masses, sfermion masses, and the gauge coupling constants,
respectively, all in the SM sector, and $m_{3/2}$ is the gravitino
mass. Notice that the absence of tree-level contribution to $m^2_\chi$ 
is guaranteed in our setup thanks to the separation of $\chi$ from
SUSY breaking. Moreover the singlet $\chi$ receives only the soft
scalar mass squared proportional to $g_X^4$. Therefore with a
sufficiently small value of the gauge coupling $g_X$, $m_\chi^2$ is
negligibly small which in turn implies that the $D$-term contribution
is suppressed.\footnote{In the anomaly mediation scenario, there is a
case where $D$-term contribution is offset by threshold corrections at
leading order~\cite{KSS}.} 
Actually enough suppression of $D$-term flavor violations requires a
condition roughly estimated as
\begin{equation}
  g_X^4 \;\lsim\; \epsilon\,g_{\rm SM}^4.
  \label{EDgb}
\end{equation}
The factor $\epsilon$ denotes an order of tuning in the sfermion
masses in order to satisfy the constraints from flavor physics
experiments. For example, a limit from the $K^0$-$\bar K^0$ mixing
phenomena implies $\epsilon\sim 10^{-(2-3)}$. This constraint is
satisfied with a natural value of $g_X$ comparable to $g_{\rm SM}$.

Another interesting pattern of spectrum is obtained by assuming that 
the SM gauge multiplets reside in the five-dimensional bulk and other 
fields are stuck on the visible brane. This situation (for the SM
sector) corresponds to gaugino-mediated supersymmetry
breaking~\cite{gauM}. The sparticle spectrum besides $D$-term
contributions is given by
\begin{eqnarray}
\begin{array}{ccl} 
  M_{\lambda_{\rm SM}} \;\sim\;
  \displaystyle{\frac{1}{\sqrt{M_5R}}m_{3/2}}, && m_{\rm SM}^2
  \;\sim\; \displaystyle{\frac{g_{\rm SM}^2}{16\pi^2} 
    M_{\lambda_{\rm SM}}^2 \ln(M_SR) 
    +\bigg(\frac{g_{\rm SM}^2}{16\pi^2}m_{3/2}\bigg)^2}, \\[4mm] 
  M_{\lambda_X} \;\sim\; \displaystyle{\frac{g_X^2}{16\pi^2}m_{3/2}},
  && m_\chi^2 \;\sim\; 
  \bigg(\displaystyle{\frac{g_X^2}{16\pi^2}m_{3/2}}\bigg)^2,
\end{array}
\end{eqnarray}
where $M_5$, $M_S$ and $R$ are the five-dimensional Planck mass,
SUSY-breaking scale, and the compactification radius of the fifth
dimension, respectively. The spectrum of the $U(1)_X$ sector,
$M_{\lambda_X}$ and $m_\chi^2$, are the same as in (\ref{3AM}). In the 
formula for the squark and slepton masses squared $m^2_{\rm SM}$, the
first term represents gaugino-mediated contribution through RGEs
[corresponding to $m_g^2$ in Eq.~(\ref{softmass})] and the last term
is that of anomaly mediation. The both terms are
generation-independent. The anomaly mediation would be dominant if
$g_{\rm SM}^2>16\pi^2/(M_5R)\ln(M_SR)$ were satisfied. However this
inequality requires too large value of $R$ even when one assumes the
five-dimensional theory is strongly coupled. In other words, it
requires unrealistically small $g_{\rm SM}$. Therefore for the SM
gauge sector, the gaugino-mediated contribution is expected to be
always larger than that of anomaly mediation. As a result, we find
that flavor violation induced by the $D$-term contribution is
suppressed if
\begin{equation}
  g_X^4 \;\lsim\; \epsilon\, \frac{16\pi^2g_{\rm SM}^2}{M_5R}\ln(M_SR).
\end{equation}
This can be easily fulfilled with an $O(1)$ gauge coupling $g_X$,
provided $g_{\rm SM}$ is properly reproduced following from the naive
dimensional analysis~\cite{NDA}.

In this way, we conclude that $D$-term contribution can indeed be
suppressed in the higher-dimensional framework without conflicting
with other phenomenological requirements such as FCNC constraints and
naturalness. Other field configurations or SUSY-breaking mechanisms in
higher dimensions could also lead to a similar conclusion.

\subsection{Suppression via superconformal dynamics}

Recently there has been proposed a new approach for realizing
hierarchical Yukawa matrices by coupling the SM sector to
superconformal (SC) gauge theories. See Ref.~\cite{NS1} for the
original proposal and also Ref.~\cite{KNT} for another. The SC sectors
are strongly coupled and assumed to have infrared fixed
points~\cite{irfp}. In this type of scenarios, the SM matter fields
couple to the SC sectors and gain large anomalous dimensions through
the strong SC dynamics. As a result, their Yukawa couplings are
suppressed at the scale where the SC sectors decouple from the SM
sector.

The SC fixed points also have an interesting implication for SUSY
breaking. The infrared convergence property of the fixed points
implies that SUSY-breaking scalar masses are also suppressed towards
the infrared~\cite{SCsoft}.\footnote{This kind of behavior was first
noticed in softly-broken SUSY QCD~\cite{SQCD}.}  
Below the SC decoupling scale, scalar masses receive radiative
corrections due to soft gaugino masses. If this mechanism is applied
to squarks and sleptons, their masses are expected to be approximately 
degenerate at low energy. Such possibility was mentioned in
Ref.~\cite{NS1} and detailed studies have been
done~\cite{SCsoft,KNNT}, where it was argued that a certain amount of
degeneracy in sfermion masses can be obtained. (See also
Ref.~\cite{LS} for a different SC approach.)

Let us discuss the $D$-term contribution in the SC framework. As
stated before, the point to avoid $D$-term flavor violation is to keep
the $U(1)_X$-breaking sector away from the SUSY-breaking one. This can
be accomplished if SC dynamics couples to the $\chi$ field whose VEV
breaks the flavor symmetry and generates flavor-dependent sfermion
masses. Then strongly-coupled SC dynamics suppresses the soft mass of
$\chi$ and consequently the $D$-term contribution to charged sfermion
masses. In the following, we suppose for simplicity that the SC sector
does not couple to the SM fields; the hierarchical structure of Yukawa
couplings are generated by the conventional FN mechanism.

Now, we present a toy model by introducing a SC gauge group 
$G_{\rm SC}$ in addition to the SM gauge and $U(1)_X$ flavor
symmetries. Consider the following cubic superpotential term
\begin{equation}
  W \;=\; h \chi\Phi\bar\Phi,
  \label{W}
\end{equation}
where $\Phi$ and $\bar \Phi$ belong to non-trivial representations
under $G_{\rm SC}$. The SM fields are $G_{\rm SC}$-singlets while
$\chi$ couples to the SC sector through the above $W$. The operator
$\Phi\bar\Phi$ has a nonzero $U(1)_X$ charge so that (\ref{W}) is
gauge invariant. Let us suppose that as in Section~2, the SM-singlet
field $\chi$ develops a non-vanishing VEV $\langle\chi\rangle$ which
breaks the flavor symmetry and generates effective Yukawa couplings
through non-renormalizable operators. An interesting point of the
present setup is that the required decoupling of the SC sector is
achieved by the same VEV $\langle\chi\rangle$, giving a mass term for
$\Phi$ and $\bar\Phi$. Thus both the SC and $U(1)_X$ sectors decouple
from the SM one around the scale of $M_X$ for $h=O(1)$.

Strong $G_{\rm SC}$ interactions force the trilinear coupling $h$ as
well as the $G_{\rm SC}$ gauge coupling to approach their infrared
fixed-point values. Hence the field $\chi$ gains a large and positive
anomalous dimension~$\gamma_\chi$. For simplicity, we assume that the
couplings at a cutoff scale $\Lambda$ are close to their fixed-point
values. If we neglect perturbatively small couplings in the SM sector,
SUSY-breaking scalar masses are suppressed (near the SC fixed point)
according to the following form of evolution equations~\cite{SCsoft}
\begin{equation}
  \mu\frac{dm^2_a}{d\mu} \;=\; {\cal M}_{ab}\,m^2_b,
  \label{ee}
\end{equation}
where the index $b$ runs over all fields coupled to the SC sector
including the $\chi$ field. The matrix ${\cal M}_{ab}$ can be
calculated by use of the Grassmannian expansion
method~\cite{softbeta}, given the anomalous dimensions $\gamma_a$ as
the functions of coupling constants in the model. Since 
${\cal M}_{ab}$ is determined by the first derivative of $\gamma_a$,
the infrared convergence property of the fixed points guarantees that
this matrix is positive definite. It follows that certain combinations
of scalar masses rapidly become suppressed through the RG evolution
from $\Lambda$ down to the the SC-decoupling scale
$\langle\chi\rangle$. Specifically, the scalar mass squared $m^2_\chi$
of the $\chi$ field behaves like
\begin{equation}
  m^2_\chi(\langle\chi\rangle) \;\simeq\; \bigg(
  \frac{\langle\chi\rangle}{\Lambda}\bigg)^{\Gamma_\chi}
  m^2_\chi(\Lambda), \qquad
  \Gamma_\chi\;\approx\;\gamma_\chi\;=\;O(1).
  \label{mchi}
\end{equation}
Here the damping factor $\Gamma_\chi$ is related to the eigenvalues
of ${\cal M}_{ab}$ and is as large as $\gamma_\chi=O(1)$. In this way,
the required separation of the flavor-breaking sector from the
SUSY-breaking one can be realized by a strong dynamics of
four-dimensional SC gauge theory. As a result, the $D$-term
contribution can be reduced for any value of SUSY-breaking parameters
at the cutoff scale.

\medskip

It is important, however, to take into account two effects that make
the desired suppression (\ref{mchi}) insufficient. One is the
violation of SC symmetry due to the $U(1)_X$ gauge interaction (as
well as the SM gauge interactions). See Ref.~\cite{KNNT} for a
detailed discussion on this point. If we include the 1-loop correction
from the $U(1)_X$ sector, the evolution equation (\ref{ee}) is
modified into
\begin{equation}
  \mu\frac{dm^2_a}{d\mu} \;=\; {\cal M}_{ab}\,m^2_b
  -\frac{8q_a^2g_X^2}{16\pi^2}\,M^2_{\lambda_X}+\cdots.
\end{equation}
Note that a model-dependent factor proportional to 
$g_X^2{\rm Tr}(Q_Xm^2)$ can safely be dropped in discussing $D$-term
induced scalar masses, as was shown in Section~2.2. The above type of
correction makes the suppression of $m^2_\chi$ incomplete, but
obviously such RG effect is less significant if the gauge coupling
$g_X$ and soft gaugino mass $M_{\lambda_X}$ are not so large.

Another and more significant effect comes from the fact that the RG
running distance in the SC regime is finite. This effect is more
important than the previous one because the running distance is
related to the hierarchy of Yukawa couplings. In the FN models
(\ref{FNspot}), the effective Yukawa couplings $y_{ij}'$ at the
SC-decoupling scale are now suppressed, in addition to the usual
factor $(\langle\chi\rangle/\Lambda)^{n_{ij}}$, by the large
wavefunction renormalization of $\chi$ and are given by
\begin{equation}
  y_{ij}' \;\simeq\; 
  \bigg(\frac{\langle\chi\rangle}{\Lambda}\bigg)^{n_{ij}}
  \bigg(\frac{\langle\chi\rangle}{\Lambda}\bigg)^{n_{ij}
  \gamma_\chi/2}\!y_{ij} \;\equiv\; 
  \big(\bar\lambda\big)^{n_{ij}}y_{ij}.
\end{equation}
Thus the degrees of suppression for Yukawa couplings and scalar masses
are correlated to each other. In particular, it is a non-trivial issue
whether $m^2_\chi$ becomes small enough to suppress FCNC with a fixed
Yukawa hierarchy $\bar\lambda$, i.e.\ with a fixed length of the SC
region. Given a value of $\bar\lambda$, the $\chi$ scalar mass
(\ref{mchi}) is written as
\begin{equation}
  m^2_\chi(\langle\chi\rangle) \;\simeq\; 
  (\bar\lambda)^{\frac{2\Gamma_\chi}{2+\gamma_\chi}}\,
  m^2_\chi(\Lambda).
  \label{dev}
\end{equation}
The deviation (\ref{dev}) from the exact convergence to the conformal
fixed point generates the $D$-term contribution to sfermion
masses. That implies the FCNC constraint for the first and second
generation squarks
\begin{equation}
  (\bar\lambda)^{1+\frac{2\Gamma_\chi}{2+\gamma_\chi}}
  \left[\frac{m^2_\chi(\Lambda)}{m^2_{\rm ave}}\right] \;\lsim\; 
  10^{-2}\bigg(\frac{m_{\rm ave}}{500\, {\rm GeV}}\bigg),
\end{equation}
where the extra factor of $\bar\lambda^1$ arises when we diagonalize
the quark mass matrices. The averaged squark mass $m_{\rm ave}$
contains initial values of squark masses at high energy and radiative
corrections from the SM gauginos. The above constraint requires that
the damping factor $\Gamma_\chi$ has an appropriately large
value. Otherwise, the $D$ problem will still be there; a high-energy
value of $m^2_\chi$ has to be small compared to other SUSY-breaking
parameters. Similar remarks apply also to the slepton sector.

In this way, suppressing the $D$-term can in principle be achieved
within four-dimensional framework, but for the mechanism discussed in
this subsection to work in practice, it is necessary to have a
concrete model of the SC sector in which $\Gamma_\chi$ is as large as
possible whereas $\gamma_\chi$ is a reasonably small.

\section{Radiative corrections from flavor sector}

In the previous section, we have discussed the dominant tree-level
$D$-term contribution which must be suppressed for models to be
phenomenologically viable. We have illustrated that the suppression
can be achieved by use of four-dimensional SC dynamics as well as a
geometrical setup in extra-dimensional scenarios. It is, however,
important to notice that even if the $D$-term contribution is
sufficiently reduced, radiative corrections due to the $U(1)_X$
gaugino may lead to sizable non-degeneracy of sfermion masses, because
the corrections depend on their quantum numbers and then become
flavor-dependent. As shown in Section 2.2, the only relevant type of
radiative corrections comes from the $U(1)_X$ soft gaugino mass. In
this section, we study this type of corrections, supposing that
tree-level $D$-term contribution is already suppressed.

We again assume for simplicity that there is only a single field
$\chi$ which breaks the gauged $U(1)_X$ flavor symmetry. As was
discussed in Eqs.~(\ref{mD}) and (\ref{mX}), the radiative correction
to sfermion masses is given by
\begin{eqnarray}
  m^2_{X_i} \;=\; \frac{-8}{16\pi^2}(q_i^2+q_\chi^2q_i)
  \int^{M_X}_\Lambda \frac{d\mu}{\mu}\, g_X^2 M^2_{\lambda_X},
  \label{mXi}
\end{eqnarray}
where $\Lambda$ is a cutoff which we take to be the Planck scale in
the following. The first term on the right-handed side originates from
a direct effect of the $U(1)_X$ gaugino in the RGE of sfermion
masses. The second term is obtained through the $D$-term contribution
with the RGE effect on $m^2_\chi$ included. Since the scalar masses
(\ref{mXi}) are the only sources of flavor dependence, the mass
difference between the scalars with different quantum numbers $q_i$
and $q_j$ is scale invariant and found to be
\begin{eqnarray}
  m^2_i-m^2_j \;=\; \frac{2}{b_X}(q_i-q_j)(q_i+q_j+q_\chi^2) 
  \Big[M^2_{\lambda_X}(\Lambda)-M^2_{\lambda_X}(M_X)\Big].
  \label{masdif}
\end{eqnarray}
The beta-function coefficient $b_X$ for $g_X$ is determined once
$U(1)_X$ charge assignment is specified. The above mass difference can
be larger than those discussed in the previous section, that is, a
small deviation from the complete suppression of $D$-term contributions.

Now let us study low-energy consequences of the gaugino radiative
correction (\ref{masdif}). The degeneracy of sfermion masses is
estimated by a ratio of off-diagonal element of scalar mass matrix to
the averaged mass $m_{\rm ave}$
\begin{equation}
  \delta_{ij} \;\equiv\; \frac{V^*_{\;ki}m^2_kV_{kj}}{m^2_{\rm ave}},
\end{equation}
where the matrix $V$ rotates the flavor basis so that Yukawa matrix is
diagonal. In the case of flavor-independent masses ($q_i=q_j$), 
$\delta_{ij}$ vanishes due to the unitarity. In the hierarchical case, 
the degeneracy is given by
\begin{eqnarray}
  \delta_{ij} &\simeq& \lambda^{|q_i-q_j|} 
  \frac{m^2_i-m^2_j}{m^2_{\rm ave}}.
  \label{dij}
\end{eqnarray}
Since the low-energy scalar masses take the form (\ref{softmass}),
$m^2_{\rm ave}$ is given by
\begin{equation}
  m^2_{\rm ave} \;=\; m_0^2+m_G^2+m_{\rm SM}^2+(m^2_{X_i}+m^2_{X_j})/2,
\end{equation}
where $m_0^2$ is the initial value (for which we take a conservative
assumption of the universality), and $m^2_G$ and $m^2_{\rm SM}$ are
the GUT and SUSY SM contributions,
\begin{eqnarray}
  m_G^2 &=& \frac{2C^G_2(R)}{b_G}\Big[M^2_{\lambda_G}(\Lambda)
  -M^2_{\lambda_G}(M_G)\Big], \\
  m_{\rm SM}^2 &=& \sum_{a=1,2,3} \frac{2C^a_2(R)}{b^a_{\rm SM}}
  \Big[M^2_{\lambda^a_{\rm SM}}(M_G) 
  -M^2_{\lambda^a_{\rm SM}}(M_S)\Big],
\end{eqnarray}
where $b$'s are the beta function coefficients of the gauge couplings,
and $M_G$ is the GUT-breaking scale. In the following analysis, we
will take a simplifying assumption that the SM gaugino masses unify to
$M_{\lambda_{\rm G}}$ at the $M_G$ scale.

As an illustrative example, we will focus on the first and second
families and take their $U(1)_X$ charges as $q_1=3$ and $q_2=2$ (and
$q_\chi=-1$). For these first two generations, there are severe
experimental constraints on scalar masses~\cite{FCNC}. As for the
squark sector, the limits on the flavor-changing scalar masses from
the Kaon system are roughly given by
\begin{eqnarray}
  \delta^Q_{12} &\lsim& 10^{-2}\,
  \bigg(\frac{m_{\rm ave}}{500\, {\rm GeV}}\bigg), \label{kaon1} \\
  \sqrt{\delta^Q_{12}\,\delta^D_{12}} &\lsim& 10^{-3}\,
  \bigg(\frac{m_{\rm ave}}{500\, {\rm GeV}}\bigg).
  \label{kaon2}
\end{eqnarray}
The former bound (\ref{kaon1}) is milder than (\ref{kaon2}) by an
order of magnitude, but is relevant for the case motivated by recent
analyses of fermion mass matrices where the Yukawa hierarchy is
ascribed to the structure of the matter fields in 10-dimensional
representation of $SU(5)$~\cite{su5-10}. For the sleptons, the 
$\mu\to e\gamma$ process requires
\begin{equation}
  \delta^L_{12} \;\lsim\; 10^{-3}\,
  \bigg(\frac{m_{\rm ave}}{100\, {\rm GeV}}\bigg)^2.
  \label{mue}
\end{equation}
The constraint on the right-handed slepton $\delta^E_{12}$ is rather
weak unless the trilinear couplings for the sleptons are very
large. This is mainly because they interact only to the weak $U(1)$
hypercharge.

\begin{figure}[htbp]
\begin{center}
\epsfxsize=10cm \ \epsfbox{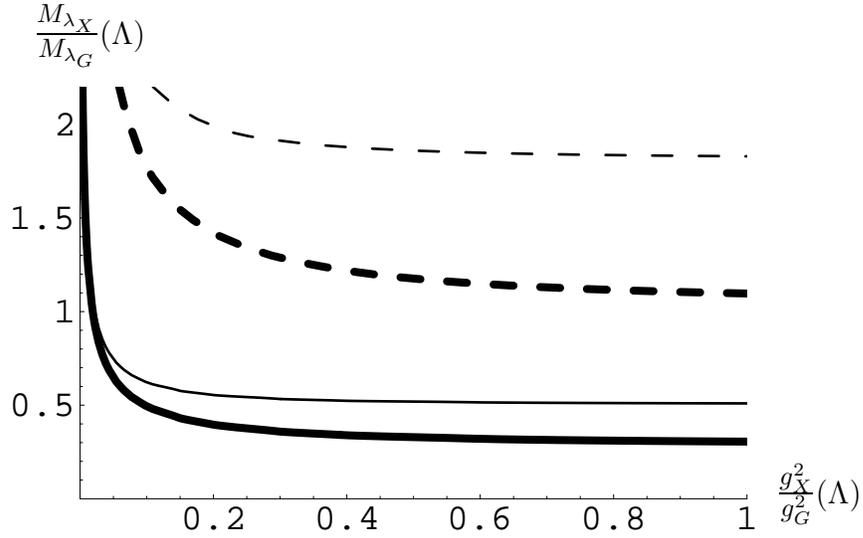}    
\put(7,16){{\large $\frac{g_X^2}{g_G^2}$}($\Lambda$)}
\put(-274,190){%
{\large $\frac{M_{\lambda_X}}{M_{\lambda_G}}$}($\Lambda$)}
\caption{The upper bounds from $K^0$-$\bar K^0$ on the $U(1)_X$ gauge
coupling and soft gaugino mass at the Planck scale. The experimentally 
allowed region is below each line. The dashed and solid lines
represent the constraints from (\ref{kaon1}) and (\ref{kaon2}),
respectively, with $b_X=100$ (bold) and $b_X=300$ (thin line). In this
figure we take $M_{\lambda_{\rm SM}}(M_G)=300$ GeV and $m_0=0$. }
\label{K}
\end{center}
\end{figure}

\begin{figure}[htbp]
\begin{center}
\epsfxsize=10cm \ \epsfbox{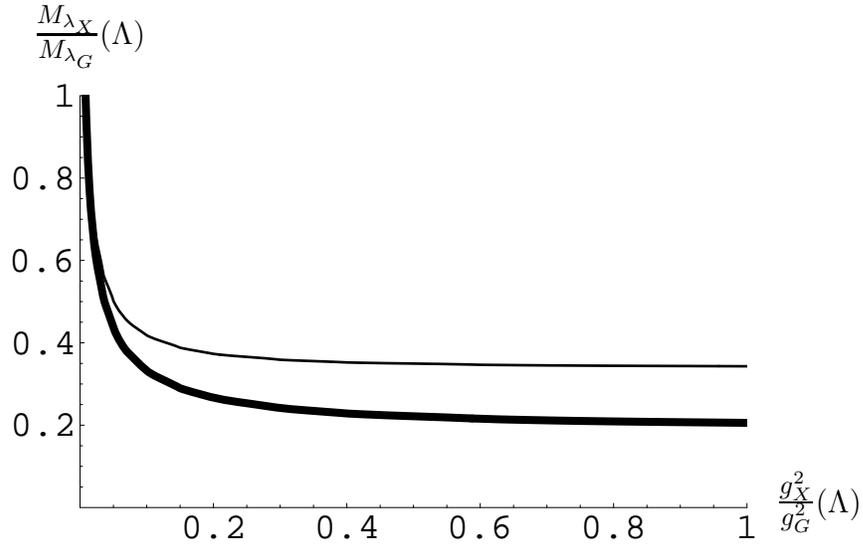}    
\put(7,13){{\large $\frac{g_X^2}{g_G^2}$}($\Lambda$)}
\put(-274,190){%
{\large $\frac{M_{\lambda_X}}{M_{\lambda_G}}$}($\Lambda$)}
\caption{The upper bounds from $\mu\to e\gamma$ on the $U(1)_X$ gauge
coupling and soft gaugino mass at the Planck scale. The others are the
same as in Figure~\ref{K}.}
\label{L}
\end{center}
\end{figure}

The above experimental upper bounds turn out to put upper bounds to
$U(1)_X$ gauge effects (the gauge coupling $g_X$ and the gaugino mass
$M_{\lambda_X}$). Assuming the minimal supersymmetric SM below the GUT
scale, there remain only a few free parameters in the mass difference
(\ref{dij}). Figure~\ref{K} shows typical upper bounds on the
parameters in the $U(1)_X$ sector.
The vertical axis denotes the ratio of $M_{\lambda_X}$ to the GUT
gaugino soft mass estimated at the Planck scale, and the horizontal
one represents $g_X^2/g_G^2(\Lambda)$. The left-lower regions below
the lines are allowed by the FCNC experiments (\ref{kaon1}) and
(\ref{kaon2}). In the figure, we take as an input the unified gaugino
mass $M_{\lambda^a_{\rm SM}}(M_G)=300$ GeV, which is suitable for the
naturalness criterion and experimental lower bounds of photino and
slepton masses. Effects of taking other values of 
$M_{\lambda_{\rm SM}}$ almost cancel in the ratios $\delta$'s and only
change the overall scale of $m_{\rm ave}$ on the right-handed sides of
(\ref{kaon1})--(\ref{mue}). In Fig.~\ref{L}, we also plot similar
experimental constraints derived from (\ref{mue}).
These two figures show that there are non-trivial limits on the
parameters of the $U(1)_X$ sector; the experimental results rule out 
large values of the gauge coupling and soft gaugino mass compared to those
of the SM sector. The $U(1)_X$ soft gaugino mass has an upper bound
which is universal in a relatively wide range of $g_X$, except for a
small $g_X$ case. For example, weakly-coupled heterotic string theory
predicts equal gaugino soft masses $M_{\lambda_G}=M_{\lambda_X}$ at
high energy, and then a tiny value of $g_X$ is required.

\begin{figure}[htbp]
\begin{center}
\epsfxsize=10cm \ \epsfbox{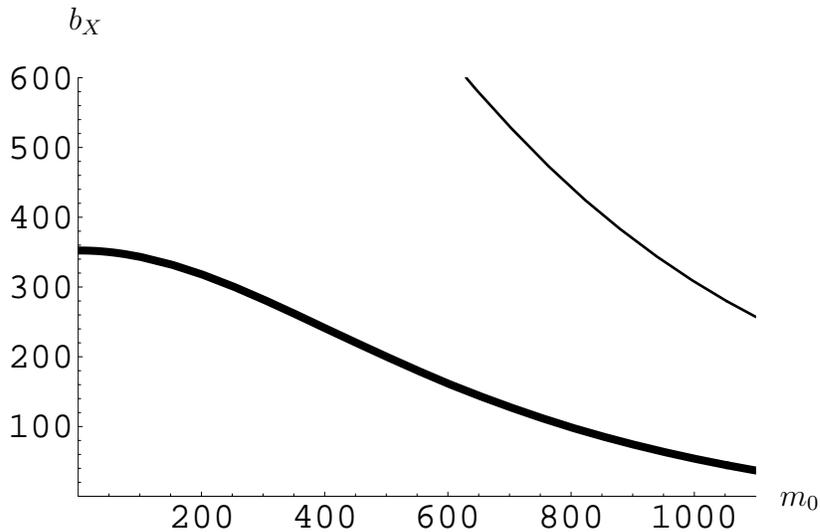}    
\put(9,10){$m_0$}
\put(-260,190){$b_X$}
\caption{The lower bounds of $b_X$ from the $K^0$-$\bar K^0$ constraint
(\ref{kaon2}). The horizontal axis denotes the initial universal
scalar soft mass $m_0$. The bold (thin) line correspond to the case 
$g_G^2(\Lambda)/4\pi^2=1/20$ (1/10). We assume
$M_{\lambda_G}(M_G)=300$ GeV and the boundary conditions $g_G=g_X$ and
$M_{\lambda_G}=M_{\lambda_X}$ at the Planck scale.}
\label{K2}
\end{center}
\end{figure}

\begin{figure}[htbp]
\begin{center}
\epsfxsize=10cm \ \epsfbox{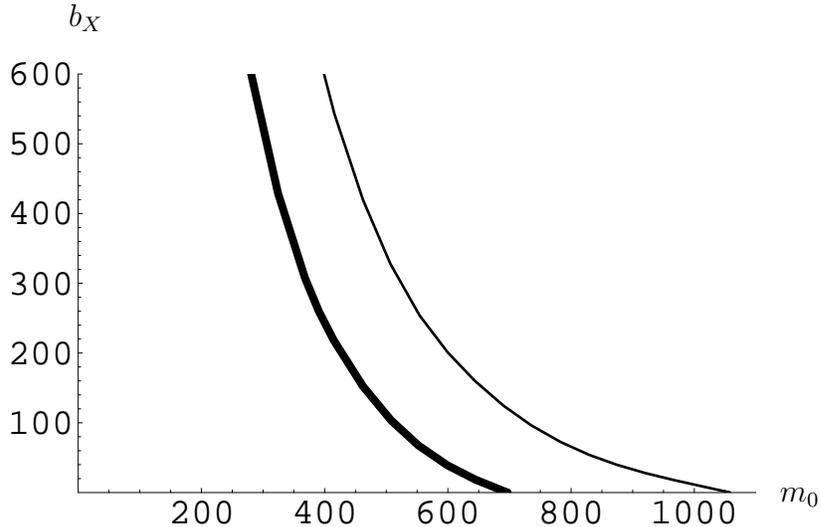}    
\put(9,10){$m_0$}
\put(-260,190){$b_X$}
\caption{Similar lower bounds of $b_X$ from the $\mu\to e\gamma$
constraint (\ref{mue}).}
\label{L2}
\end{center}
\end{figure}

Given fixed values of gauge couplings and gaugino masses, the FCNC
constraints give a lower bound for the $U(1)_X$ gauge beta function
$b_X$. Note that models with abelian flavor symmetry generally contain
a number of charged fields to realize large Yukawa hierarchy (and also
to stabilize GUT Higgs potential). The $U(1)_X$ gauge beta-function
can therefore be relatively large. In Figs.~\ref{K2} and~\ref{L2}, we
show typical lower bounds for $b_X$ with the unification boundary
conditions $g_G=g_X$ and $M_{\lambda_G}=M_{\lambda_X}$ at the Planck
scale.
The two types of lines in the figures represent the cases
$g_G^2(\Lambda)/4\pi=1/20$ (bold) and 1/20 (thin), respectively. The
horizontal axis denotes the initial universal scalar soft mass $m_0$
which is given at the Planck scale by some SUSY-breaking dynamics and
then could not be much larger than gaugino masses. For the sleptons 
(Fig.~\ref{L2}), introducing a nonzero $m_0$ considerably changes the 
$b_X$ bounds than in the squark sector (Fig.~\ref{K2}). We find from
these figures that for a fixed model (a fixed $b_X$), a rather small
value of $g_X$ and/or a large initial scalar mass $m_0$ must be
satisfied. The constraints derived in this section generically exist
and are independent of mechanisms for avoiding flavor-violating
tree-level $D$ terms. Therefore they could cast a serious problem on
the construction of realistic flavor $U(1)_X$ models from high-energy
fundamental theories.

\medskip

In Section 3, we have found that the dominant $D$-term contribution
can be suppressed by use of particular setups if the $U(1)_X$ gauge
coupling satisfies a weak condition like (\ref{EDgb}). The bounds
discussed here may be stronger than those obtained in Section
3. Furthermore the bounds from the radiative corrections are roughly
model-independent and can be applied to any dynamics suppressing
$D$-term contribution. For example, in the extra-dimensional models in
Section 3.2, the $U(1)_X$ vector multiplet is taken to be stuck on the
visible brane. Then a suppressed initial value of $U(1)_X$ soft
gaugino mass is obtained because of the absence of local operators
with which the gaugino directly couples to the SUSY-breaking
brane. For the superconformal scenarios in Section 3.3, the
suppression of $M_{\lambda_X}$ cannot naively be achieved due to the
fact that an abelian factor does not have any infrared fixed point of
the gauge coupling. Embedding it to a non-abelian gauge group may cure
the problem, but in that case it might be non-trivial to construct
realistic Yukawa hierarchy.

Our analysis here shows that flavor-dependent radiative corrections
due to the $U(1)_X$ gaugino is important from a viewpoint of
experimental constraints on flavor physics. In this paper we have only
discussed flavor-violating scalar masses induced by gauged abelian
flavor symmetries. In addition to this, CP-violation phenomena, RGE
effects on scalar trilinear couplings, etc.\ would give severer bounds
on gauged flavor symmetries. Those issues will be discussed elsewhere.

\section{Summary}

We have studied sfermion masses in supersymmetric models with gauged
$U(1)$ flavor symmetries. It has been known that these models
generally suffer from large flavor-violating effects due to tree-level
sfermion masses proportional to individual $U(1)$ quantum numbers. We
have presented a simple idea for suppressing the $D$-term
contribution; a separation of flavor-symmetry breaking from
supersymmetry breaking. Illustrative models have been described by
extra-dimensional frameworks or strongly-coupled superconformal dynamics.

We have also pointed out that generation-dependent radiative
corrections due to the $U(1)$ soft gaugino mass lead to sizable
non-degeneracy in sfermion masses. The corrections are also dangerous
to the FCNC constraints and put a non-trivial limit on the parameters
in the flavor symmetry sector. Our analysis indicates that it is
preferable to have the $U(1)$ gauge coupling and soft gaugino mass as
small as possible. Required small values of parameters could be
obtained, for example in extra-dimensional models, by volume
suppression factors or strong RGE effects.

We add a remark that none of the flavor-violating effects is present
if one could take the global limit $g\to 0$. However if one takes
serious the argument that any global symmetry is not preserved in the
presence of quantum gravity, one must seek for solutions to avoid the
flavor-violating $U(1)$ gauge contributions. Moreover, it seems
difficult to experimentally test such models with global flavor
symmetry. On the other hand, the gauged flavor symmetry will provide
us with detectable signatures especially through the flavor-dependent
radiative corrections from $U(1)$ gaugino.

\subsection*{Acknowledgments}

The authors thank the Yukawa Institute for Theoretical Physics at
Kyoto University, where a portion of this work was carried out during
the YITP-W-02-02 on "Progress in Particle Physics". T.~K.\ and K.~Y.\
are supported in part by the Grants-in-Aid for Scientific Research
No.~14540256 and the Grant-in-Aid No.~07864, respectively, from the
Ministry of Education, Science, Sports and Culture, Japan.


\end{document}